\pdfoutput=1	

\documentclass[osajnl,twocolumn,showpacs,10pt]{revtex4-1}
\usepackage{amsmath,amssymb,graphicx}
\begin{document}

\title{Perfect imaging with planar interfaces}

\author{Stephen Oxburgh}
\author{Johannes Courtial}
\email{Corresponding author: Johannes.Courtial@glasgow.ac.uk}
\affiliation{School of Physics \& Astronomy, University of Glasgow, Glasgow G12~8QQ, United~Kingdom}

\begin{abstract}
We describe the most general homogenous, planar, light-ray-direction-changing sheet that performs one-to-one imaging between object space and image space. This is a non-trivial special case (of the sheet being homogenous) of an earlier result [J.\ Courtial, Opt.\ Commun.\ \textbf{282}, 2480 (2009)]. Such a sheet can be realised, approximately, with generalised confocal lenslet arrays.
\end{abstract}


\maketitle 

\section{Introduction}

\noindent
Much work has been done in recent years designing optical devices which exhibit ray-optically perfect imaging.
Historically the first such devices were the Maxwell fisheye \cite{Maxwell-1853,Maxwell-1854} and the Luneburg \cite{Luneburg-1944} and Eaton \cite{Eaton-1952} lenses, all of which were recently described as special cases of a more general class of ``lens'' \cite{Tyc-et-al-2011, Sarbort-Tyc-2012}.
Superlenses \cite{Pendry-2000,Fang-et-al-2005} and hyperlenses \cite{Jacob-et-al-2006,Liu-et-al-2007} are also producing ray-optically perfect images (and --- remarkably --- the images are also wave-optically perfect, as are those in the Maxwell fisheye \cite{Leonhardt-2009}).
Less obviously, this class also includes devices such as invisibility cloaks \cite{Leonhardt-2006,Pendry-et-al-2006}, which are invisible because they image each point to itself.
A number of wave-optical theorems relate to perfect imaging~\cite{Tyc-et-al-2011}.

A planar surface that refracts according to the generalised law of refraction $\tan \alpha_1 = \eta \tan \alpha_2$, where $\alpha_1$ and $\alpha_2$ are the angles of incidence and refraction, respectively, and $\eta$ is a constant, also performs perfect imaging: the image position is the same as the object position, but its distance from the surface is multiplied by a factor $\eta$ \cite{Courtial-2008a}.
This direction change can be realised with an array of miniaturised telescopes, built from confocal lenslet arrays (CLAs). CLAs are composed of two arrays of microlenses which share a common focal plane which form small telescoplets. However CLAs introduce an imperfection in the form of a small (and often negligible) offset of the light-ray position~\cite{Courtial-2008a}.
Wave-optically, the ray offset is due to the optical elements introducing something similar to the pixelation of a computer monitor:
close inspection reveals that each optical element transforms one piece of the beam, but when viewed from a sufficiently great distance, the pieces cannot be resolved and the beam transformation appears to be point by point.
Throughout this paper, we ignore this offset/pixellation, as it can be made negligible in practice.
CLAs are examples of METATOYs \cite{Hamilton-Courtial-2009}, which are arrays of optical elements larger than the wavelength of light, such as telescopes, and provided the elements work across the entire visible spectrum, so do the METATOYs.
The light-ray fields behind METATOYs can appear to (but not actually) be wave-optically forbidden~\cite{Hamilton-Courtial-2009}.

\begin{figure}
\begin{center} \includegraphics{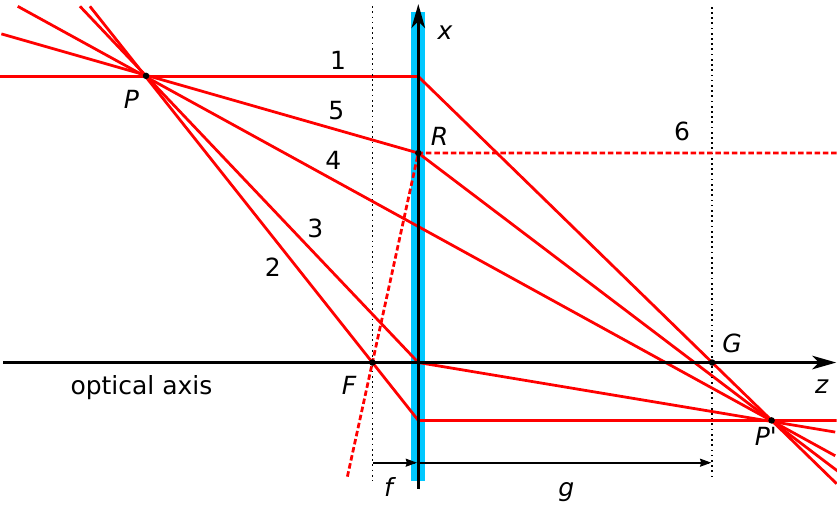} \end{center}
\caption{\label{imaging-sheet-figure}Light-ray trajectories (red lines) through an imaging sheet (thick, vertical, turquoise line).
The imaging sheet refracts like a combination of an idealised thin lens of focal length $f$ and CLAs \cite{Courtial-2008a} (which strain image space by a factor $\eta$ in the axial direction).
$F$ and $G$ are the object- and image-sided focal points, located respectively a distance $f$ in front of the sheet and a distance $g = \eta f$ behind it.
The solid rays are involved in imaging $P$ to $P'$; rays 1 to 3 are the principal rays through the object- and image-sided focal points and through the intersection between the sheet and the optical axis, respectively.
Note that ray 3  does not pass straight through the sheet, but other rays do (for example ray 4).
Apart from passing through $P$ and $P'$, ray 5 is not special in any way; we call the point where it intersects the sheet $R$.
Ray 6 (dashed) is the principal ray through $R$ that passes through the object-sided focal point and is parallel to the optical axis in image space.}
 \end{figure}

A previous paper \cite{Courtial-2009b} considered the most general type of inhomogeneous,
infinite, planar, non-absorbing, light-ray-direction-changing sheet that can perform perfect one-to-one imaging of any point in object space into a corresponding point in image space (and vice versa); here we call such a sheet an \emph{imaging sheet}.
Ref.\ \cite{Courtial-2009b} showed that an imaging sheet is a combination of an idealised thin lens and (zero-offset) CLAs (Fig.\ \ref{imaging-sheet-figure}).
The considerations were purely ray-optical in nature, and therefore not subject to any limitations due to wave-optical theorems about perfect imaging.
In fact, CLAs are components that refract according to a generalised law of refraction that can lead to wave-optically forbidden light-ray fields \cite{Courtial-Tyc-2012}.
Here we return to this earlier work and explore in greater detail the key results in the limit where the imaging sheet becomes homogenous.
In doing so we show that a subset of generalised CLAs (gCLAs) \cite{Hamilton-Courtial-2009b}, again in the limit of zero ray offset, are examples of such sheets.

\section{Review of imaging with inhomogeneous sheets}

\noindent
We expand on the work already presented in Ref.\ \cite{Courtial-2009b}.
We briefly state the necessary assumptions:
\begin{enumerate}

\item The sheet we consider is infinitely thin, infinite, planar and non-absorbing and changes the direction of the transmitted light rays without offsetting the ray position (i.e.\ the ray leaves the sheet from the same point where it strikes).

\item The sheet is placed in a homogenous medium and as such the light rays on both sides of the sheet travel in straight lines.

\item The sheet performs one-to-one imaging between \emph{all} of object space and \emph{all} of image space.

\end{enumerate}
In the remainder of this paper we call such a sheet an imaging sheet.

The following relations were found between object and image positions for an imaging sheet in the $z=0$ plane:
\begin{equation}
\label{object-image-relations-1}
\frac{x'}{x}=\frac{f}{f-o}, \quad
\frac{y'}{y}=\frac{f}{f-o}, \quad
\frac{f}{o}+\frac{g}{i}=1,
\end{equation}
where $x$, $y$ and $x'$, $y'$ are the $x$- and $y$-components of the object and image points respectively, $o$ is the object distance, $i$ is the image distance, and $f$ and $g$ are arbitrary constants.
The object and image distance respectively becomes $o = -z$ and $i = z'$, where $z$ is the $z$ coordinate of the object position and $z'$ is the $z$ coordinate of the image position.
We can then write the relations between object and image positions as
\begin{equation}
\label{object-image-relations-2}
\frac{x'}{x}=\frac{f}{f+z}, \quad
\frac{y'}{y}=\frac{f}{f+z}, \quad
\frac{z'}{z} = \frac{g}{f}.
\end{equation}

In the form (\ref{object-image-relations-1}), these equations are similar to the equations describing lens imaging, but instead of one focal length the lens now has two different focal lengths for object and image space, namely $f$ and $g$.
Such a lens is equivalent to an idealised thin lens with (object- and image-sided) focal length $f$ and CLAs placed on one side such that image space is additionally strained in the direction of the optical axis by a factor $\eta$, and with it the image-sided focal length, which becomes $g = \eta f$ (Fig.\ \ref{imaging-sheet-figure}).

It is perhaps worth noting that imaging sheets cannot be realised in the form of thin phase holograms.
This can be seen from Fig.\ \ref{imaging-sheet-figure} as follows.
The light-ray-direction change on transmission through a phase hologram is due to the transverse phase gradient the hologram adds to the ray.
Assume that the sheet can be realised in the form of a phase hologram.
Then the fact that ray 4 passes through the sheet undeviated suggests that this hologram imparts zero additional transverse phase gradient at the point where ray 4 passes through it.
However, other light rays that pass through the same point do change direction; for example, the light ray that first passes through the object-sided focal point, $F$, would be re-directed so that it is parallel to the optical axis after transmission through the sheet.
This means that the assumption that a phase hologram can act as an imaging sheet leads to inconsistencies, and is therefore wrong.

\section{\label{homogeneous-sheet-section}The special case of homogeneous sheets}

\begin{figure}
\begin{center} \includegraphics{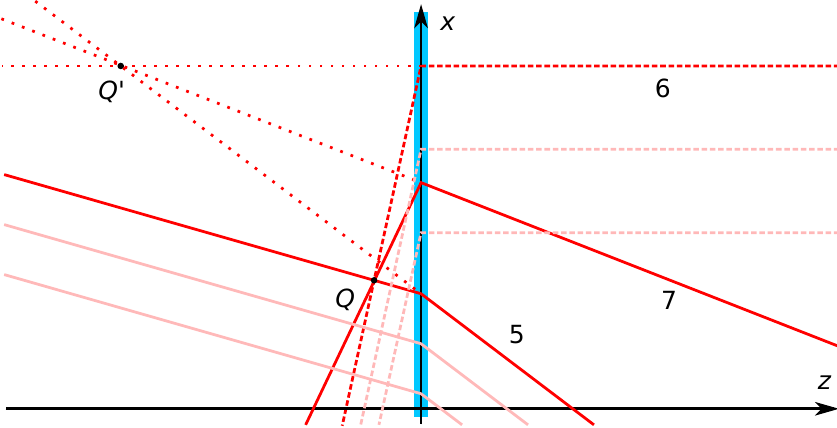} \end{center}
\caption{\label{homogeneous-imaging-sheet-figure}Light-ray trajectories (red lines) through a homogeneous imaging sheet (thick, vertical, turquoise line).
In this example, the law of refraction from a point, $R$, on the (inhomogeneous) imaging sheet shown in Fig.\ \ref{imaging-sheet-figure} applies across the entire sheet.
Two rays, marked 5 and 6, with the same incident and outgoing directions as the corresponding rays through $R$ shown in Fig.\ \ref{imaging-sheet-figure} are used to construct the position of the image of a point $Q$, $Q'$.
Ray 7 is an example of an additional light ray that passes through $Q$.
As the sheet images all points, after transmission through the sheet the ray (or its straight-line continuation) passes through $Q'$.}
 \end{figure}

\noindent
We start from the description of the general mapping performed by an imaging sheet as a combination of an idealised thin lens and CLAs (Fig.\ \ref{imaging-sheet-figure}).
The special case of the imaging sheet being homogeneous has to be describable like this, too.
The CLAs are already homogeneous, so the case of homogeneous imaging sheets has to correspond to the case where the lens becomes homogeneous, which happens for $f \rightarrow \infty$.

To construct the general limit $f \rightarrow \infty$, imagine isotropically scaling up the size of the diagram in Fig.\ \ref{imaging-sheet-figure} around point $R$.
This means increasing the image- and object-sided focal lengths and the distance between the optical axis and $R$, all by the same factor.
As the scaling factor is increased, the part of the diagram around $R$ becomes increasingly homogeneous.
In the limit of the scaling factor reaching infinity, the imaging sheet \textit{is} homogeneous, and the law of refraction at point $R$ on the initial, inhomogeneous, imaging sheet applies across the entire sheet (Fig.\ \ref{homogeneous-imaging-sheet-figure}).

We derive the mapping between object and image positions in this limit as follows.
First, we define new coordinates relative to the coordinates of $R$ as follows:
\begin{equation}
\begin{aligned}
\tilde{x} &= x - R_x, \quad
\tilde{y} = y - R_y, \quad
\tilde{z} = z, \\
\tilde{x}' &= x' - R_x, \quad
\tilde{y}' = y' - R_y, \quad
\tilde{z}' = z',
\end{aligned}
\label{relative-coordinates-definition}
\end{equation}
where $R_x$ and $R_y$ are the $x$ and $y$ coordinates of $R$.
In terms of these new coordinates, we re-write the first two equations in Eqns (\ref{object-image-relations-2}) to get
\begin{equation}
\frac{R_x+\tilde{x}'}{R_x+\tilde{x}} = \frac{f}{f+\tilde{z}}, \quad
\frac{R_y+\tilde{y}'}{R_y+\tilde{y}} = \frac{f}{f+\tilde{z}}.
\label{object-image-relations-3}
\end{equation}

Next, we introduce the parameters
\begin{equation}
t_x = \frac{R_x}{f}, \quad
t_y = \frac{R_y}{f}.
\label{t-equations}
\end{equation}
Rearranging the first equation in Eqns (\ref{object-image-relations-3}) and expressing the result terms of these new parameters gives
\begin{equation}
\begin{aligned}
\label{x1-primed-equation}
\tilde{x}'
&= \frac{f}{f+\tilde{z}}(f t_x + \tilde{x}) - f t_x
= \frac{f}{f+\tilde{z}} (\tilde{x} - \tilde{z} t_x ).
\end{aligned}
\end{equation}
In the limit when $f\rightarrow\infty$, this reduces to
\begin{equation}
\tilde{x}' = \tilde{x} - \tilde{z} t_x.
\label{x'-equation}
\end{equation}
Similarly,
\begin{equation}
\tilde{y}' = \tilde{y} - \tilde{z} t_y.
\label{y'-equation}
\end{equation}

The third equation in (\ref{object-image-relations-2}), after substituting $\eta f$ for $g$ and dividing by $f$, becomes in the limit $f \rightarrow \infty$
\begin{equation}
\tilde{z}' = \eta \tilde{z},
\label{z'-equation}
\end{equation}
where $\tilde{z}$ and $\tilde{z}'$ are the longitudinal coordinates of the object position and image position, respectively.
Finally, we drop the tilde from the coordinate names, which gives the following equations describing the mapping between object and image space:
\begin{equation}
x' = x - z t_x, \quad
y' = y - z t_y, \quad
z' = \eta z.
\label{mapping-equations}
\end{equation}
This is the main result of this paper.
It can alternatively be written in the following vector form:
\begin{equation}
\bm{P}'
= \bm{P} - z \left( \begin{array}{c} t_x \\ t_y \\ 1 - \eta \end{array} \right),
\end{equation}
where $\bm{P} = (x, y, z)^T$ is a vector to the object position and $\bm{P}' = (x', y', z')^T$ is a vector to the image position.

\section{gCLAs as homogenous imaging sheets}

\noindent
It would be desirable to have a component that images like the homogeneous imaging sheet described above.
To do so, the component would have to change light-ray direction in the same way as the homogeneous imaging sheet.

Luckily, the light-ray-direction change for homogeneous imaging sheets can be realised with a subset of generalised confocal lenslet arrays (gCLAs) \cite{Hamilton-Courtial-2009b}.
As the name suggests, gCLAs are generalised CLAs \cite{Courtial-2008a}, which, as we already mentioned in the \emph{Introduction}, are arrays of telescopelets.
The parameter $\eta$ hat characterises CLAs is given by $-f_2 / f_1$, where $f_1$ and $f_2$ are respectively the focal length of the first and second lens of each telescopelet.
In CLAs, each telescopelet is orientated such that its optical axis is perpendicular to the (tangent) plane of the CLAs.
In gCLAs, each telescopelet is generalised in a number of ways, such that the gCLAs retain their METATOY characteristic of exhibiting homogenous refraction, which means that a light ray with a given ingoing direction can strike any point on the sheet and it will be refracted with the same outgoing direction. Like CLAs, gCLAs introduce between the ingoing and outgoing light rays an offset, which is usually negligible.
Like in CLAs, light rays can also enter through lens 1 of one telescopelet and exit through lens 2 of another telescopelet, which leads to additional images \cite{Courtial-2009,Maceina-et-al-2011}.

\begin{figure}
\begin{center} \includegraphics{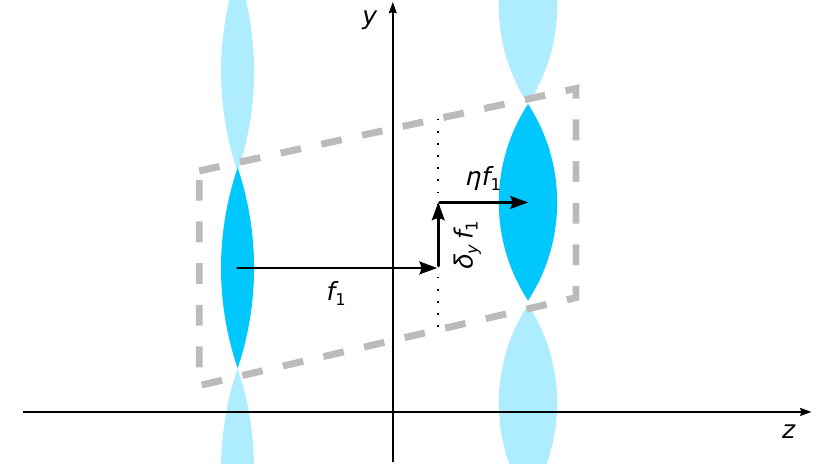} \end{center}
\caption{\label{gCLAs-figure}Fine structure of the generalised confocal lenslet arrays (gCLAs) relevant in the context of this paper.
The gCLAs consist of arrays telescopelets, each comprising two lenslets sharing a common focal plane (dotted vertical line), forming a sheet.
One telescopelet is surrounded by a dashed grey line.
The focal length of the lenslets is $f_1$ and $f_2 = \eta f_1$, respectively.
Both optical axes of the two lenslets are perpendicular to the plane on which the sheet of gCLAs is centred (here the $z=0$ plane).
The optical axis of the second lenslet is offset with respect to that of the first by $\delta_x f_1$ in the $x$ direction and by $\delta_y f_1$ in the $y$ direction.}
\end{figure}

Only one of the generalisations of CLAs to gCLAs is relevant in the context of this paper, namely a sideways displacement of the optical axes of the lenslets that form the telescopelet relative to each other (Fig.\ \ref{gCLAs-figure}).
This sideways displacement is characterised by the dimensionless parameters $\delta_x$ and $\delta_y$, which are defined as the displacement between the optical axes in the $x$ and $y$ direction, respectively, divided by $f_1$ \cite{Hamilton-Courtial-2009b}.
(When fully generalised, the each lenslet has two focal lengths, in the case of lens 1 these would be called $f_{x1}$ and $f_{y1}$.
This generalisation is not necessary here, where we consider the case $f_{x1} = f_{y1} = f_1$.)
The law of refraction for gCLAs generalised in this way is (Eqn (11) in Ref.\ \cite{Oxburgh-Courtial-2013b}, for the case where the $\hat{\bm{u}} = \hat{\bm{x}}$, $\hat{\bm{v}} = \hat{\bm{y}}$, $\hat{\bm{a}} = \hat{\bm{z}}$, and $\eta_u = \eta_v = \eta$, after multiplication by $d_z$)
\begin{equation}
d'_x = \frac{d_x - d_z \delta_x }{\eta}, \quad
d'_y = \frac{d_y - d_z \delta_y}{\eta}, \quad
d'_z = d_z,
\label{gCLAs-law-of-refraction}
\end{equation}
where the vectors $\bm{d} = (d_x, d_y, d_z)$ and $\bm{d}' = (d'_x, d'_y, d'_z)$ are vectors in the direction of the incident and outgoing light-ray direction, respectively.
(Note that we have not multiplied by $\eta$ in Eqns (\ref{gCLAs-law-of-refraction}), as this would reverse the direction of $\bm{d}'$ if $\eta < 0$.)

\begin{figure}
\begin{center} \includegraphics{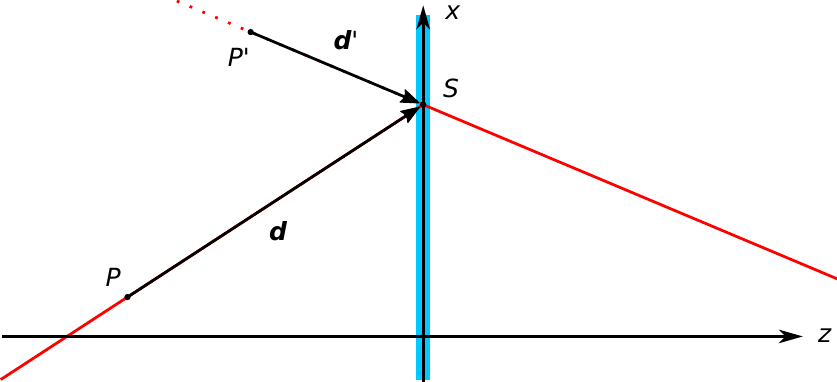} \end{center}
\caption{\label{law-of-refraction-figure}Law of refraction for a homogeneous imaging sheet.
The relationship between the vectors $\bm{d}$ and $\bm{d}'$, which point in the direction of the incident and outgoing light-ray direction, respectively, can be derived from the light-ray trajectory (dotted line) through a point $P$ and its image, $P'$, via an arbitrary point $S$ on the imaging sheet.
The diagram is drawn for $\eta > 0$, and so the image is on the same side of the sheet as the object.
The the object is drawn to be real, the image is virtual.}
\end{figure}

\begin{figure}
\begin{center}
\includegraphics[width=\columnwidth]{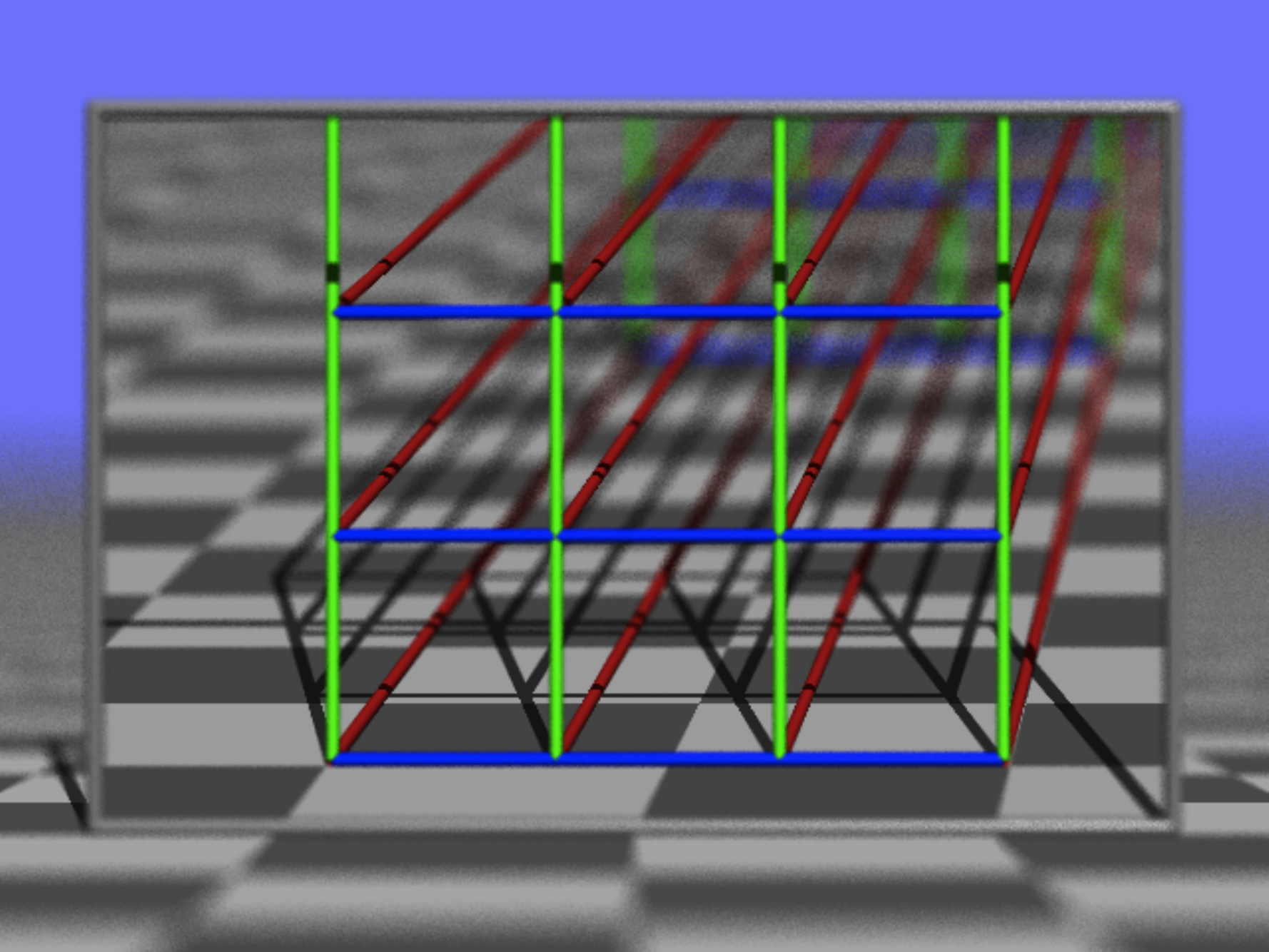} \\
$z=11$ \\
\vspace{0.5cm}
 \includegraphics[width=\columnwidth]{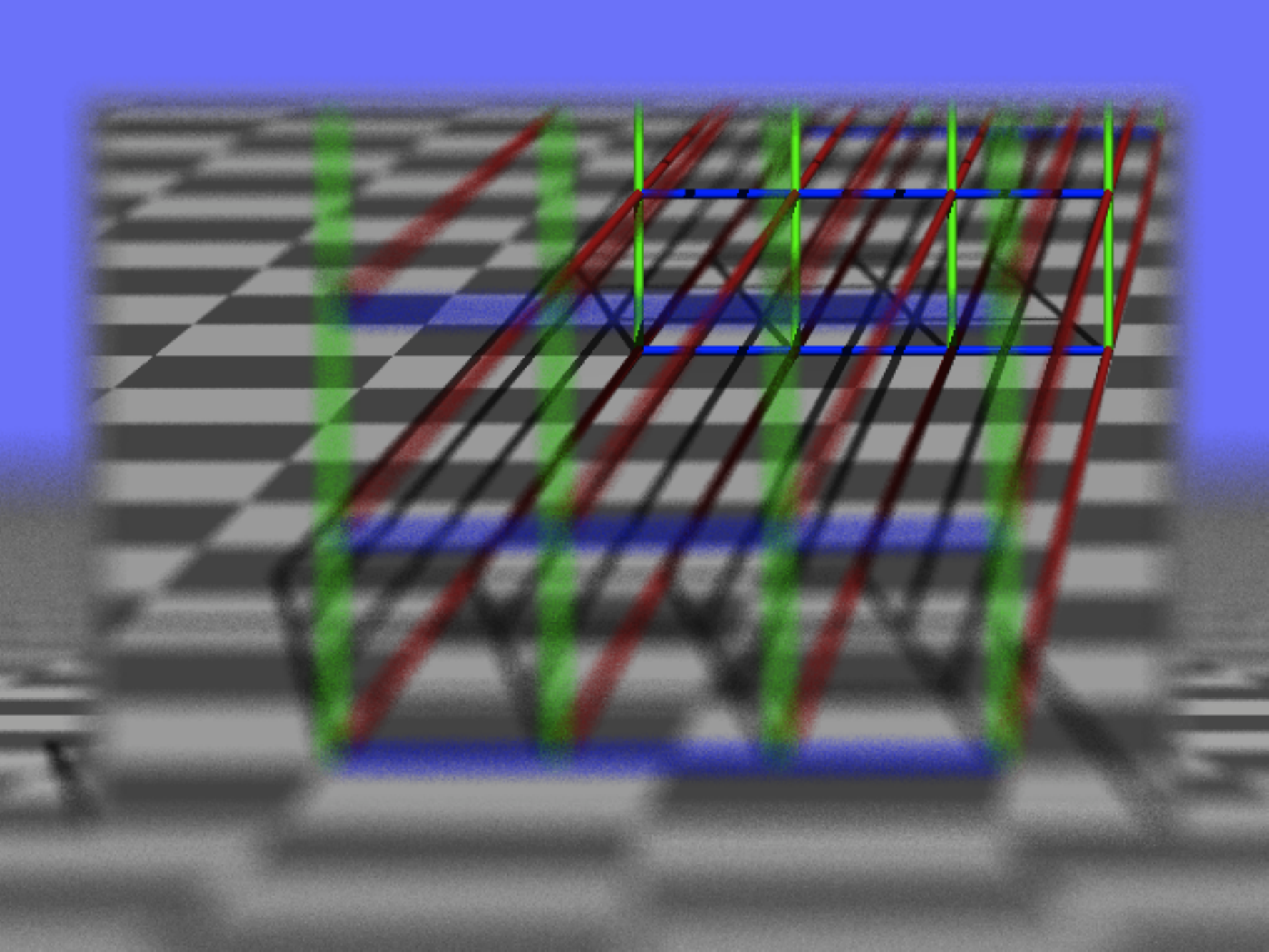} \\
$z=15$
\end{center}
\caption{\label{fig:gCLAs-focussing}Generalised confocal lenslet arrays (gCLAs) as homogeneous imaging sheets.
In these ray-tracing simulations, an extended object is seen through an example of gCLAs.
The depth of focus is relatively short, blurring any part of the image that is not in focus.
Focussing on different planes brings different parts of the object into sharp focus, demonstrating that the gCLAs image all those different parts (as they should --- as imaging sheets, they must image \emph{all} object space).
The extended object is a 3-dimensional lattice of coloured cylinders.
The parameters of the gCLAs are $\eta=2$, $\delta_x=0.2$, $\delta_x=0.3$; they are positioned in a plane a distance of 10 floor-tile lengths in front of the camera.
The focussing distance is $z$ (again in floor-tile lengths).
The simulations were performed using the open-source ray-tracing program TIM \cite{Lambert-et-al-2012,Oxburgh-et-al-2013}.}
\end{figure}

We now show that the mapping described in section \ref{homogeneous-sheet-section} corresponds to the same law of refraction.
This law of refraction, i.e.\ the relationship between the outgoing and incoming light-ray directions, can be easily established by considering a light-ray trajectory between two points that are imaged into each other by the sheet, $P$ and $P'$, via an arbitrary point $S$ on the sheet (Fig.\ \ref{law-of-refraction-figure}).
If the sheet is placed in the $z=0$ plane, the coordinates of $P'$, $(x', y', z')$, are related to the coordinates of $P$, $(x, y, z)$, according to Eqns (\ref{mapping-equations}).
We call the coordinates of $S$ $(S_x, S_y, 0)$.
The coordinates of the vector $\bm{d}$, which points in the direction of the incident light ray, can then be written as
\begin{equation}
d_x = S_x - x, \quad
d_y = S_y - y, \quad
d_z = - z.
\label{d-equations}
\end{equation}
The coordinates of the vector $\bm{d}'$, which points in the direction of the outgoing light ray, becomes
\begin{equation}
d'_x = S_x - x', \quad
d'_y = S_y - y', \quad
d'_z = -z'.
\label{dPrime-equations}
\end{equation}
Substituting Eqns (\ref{mapping-equations}) into Eqns (\ref{dPrime-equations}), and dividing by $\eta$, gives
\begin{equation}
\begin{aligned}
d'_x &\propto \frac{S_x - x + z t_x}{\eta}  = \frac{d_x - d_z t_x}{\eta}, \\
d'_y &\propto \frac{S_y - y + z t_y}{\eta} = \frac{d_y - d_z t_y}{\eta}, \\
d'_z &\propto - z = d_z.
\end{aligned}
\label{dPrime-equations-2}
\end{equation}
(The equations were derived from Fig.\ \ref{law-of-refraction-figure}, which is drawn for $\eta > 0$.  By writing them in the above form, with $\eta$ in the denominator, they are correct also for $\eta < 0$.)
Provided that
\begin{equation}
\delta_x = t_x, \quad
\delta_y = t_y,
\end{equation}
the outgoing light-ray directions $\bm{d}'$ described by Eqns (\ref{gCLAs-law-of-refraction}) and (\ref{dPrime-equations-2}) are proportional to each other.
The law of refraction for gCLAs is therefore identical to that of homogeneous imaging sheets.
The above finding means that gCLAs are approximations to homogeneous imaging sheets (which is demonstrated in Fig.\ \ref{fig:gCLAs-focussing}); the approximation is as good as the offset experienced by light rays on transmission through gCLAs is negligible.

It is worth noting that there is currently no realisation of homogeneous imaging sheets other than in terms of gCLAs.
This is possibly for fundamental reasons:  the law of refraction for homogeneous imaging sheets, Eqn (\ref{dPrime-equations-2}), can lead to wave-optically forbidden light-ray fields \cite{Courtial-Tyc-2012}, which does not necessarily imply that it is impossible to build wave-optically perfect homogeneous imaging sheets (with no offset or other imperfections), but so far only laws of refraction that \emph{never} lead to wave-optically forbidden light-ray fields (one is Snell's-law refraction, the other phase-hologram refraction \cite{Yu-et-al-2011}) have been realised such that they are wave-optically perfect~\cite{Courtial-Tyc-2012}.

\section{Conclusion}

\noindent
By building upon earlier work on inhomogeneous imaging sheets, we have established the imaging properties of homogenous imaging sheets.
We have calculated the generalised law of refraction for such sheets, and we have shown that it can be realised with generalised confocal lenslet arrays (gCLAs).

As the law of refraction for homogeneous imaging sheets can be realised with gCLAs, and as each homogeneous imaging sheets simply refract across the entire sheet like an inhomogeneous imaging sheets does at one particular point on the sheet, inhomogeneous imaging sheets can be realised with inhomogeneous gCLAs.

We conclude that gCLAs could well find use in imaging applications.


\end{document}